\documentclass[11pt]{article}

\columnwidth\textwidth
\usepackage{graphicx}
\usepackage{epsfig}

\begin{document}
\title{Coulomb Corrections for Coherent Electroproduction of
Vector Mesons: Eikonal Approximation}
\author{Andreas Aste, Kai Hencken, Dirk Trautmann\\ 
Institut f\"ur Physik der Universit\"at Basel, 4056 Basel,
Switzerland}
\date{July 8, 2004}

\maketitle

\begin{center}
\begin{abstract}
Virtual radiative corrections due to
the long range Coulomb forces of heavy nuclei with
charge $Z$ may lead to sizeable corrections to the Born cross
section usually used for lepton-nucleus scattering processes.
An introduction and presentation of the most important
issues of the eikonal approximation is given.
We present calculations for forward electroproduction
of rho mesons in a framework suggested by the VDM
(vector dominance model), using the eikonal approximation.
It turns out that Coulomb corrections may become relatively large.
Some minor errors in the literature are corrected.
Electroproduction
reactions - 13.40.-f Electromagnetic processes and properties - 25.30.Bf
Elastic electron scattering
\vskip 0.2 cm
\noindent Eur. Phys. J. A (2004)\\
\noindent DOI 10.1140/epja/i2003-10182-3
\end{abstract}
\end{center}

\newpage

\section{Introduction} 
Due to the small size of the elemental electric charge
$e=\sqrt{4 \pi \alpha}$,
it is for most electromagnetic elementary particle reactions
fully sufficient to calculate only in Born approximation.
But this is no longer true
when heavy nuclei are involved, like e.g. lead, where the relevant
perturbation expansion parameter $\alpha Z \sim 0.6$ is not small.
It is then possible that even at high energies, the ratio of the exact and
the Born cross section does not approach unity.
One prominent example for this observation
is the photoelectric effect, where a single photon
knocks out an electron from the $K$ shell, or electron positron pair
production by a single photon incident on a nucleus, where the
electron is captured into the $K$ shell \cite{Aste,Agger}.
For small $\alpha Z$
and high photon energy $\epsilon_\gamma$, the cross section
for these processes is given by the Sauter formula \cite{Sauter}
\begin{equation}
\sigma_0 \simeq 4 \pi \alpha^6 Z^5 \lambda_C^2 \frac{1}{\epsilon_\gamma} ,
\end{equation}
where $\lambda_C$ is the electron Compton wavelength.
Applying the Born approximation in the usual sense, i.e. by the
use of plane waves for the positron, was attempted by Hall and
Oppenheimer \cite{Oppenheimer} already in 1931 in the case of the
photoelectric effect.
But a central difficulty in the treatment of the photoeffect
arose from the distorted wave function of the
ejected electron. The bound state wave function depends on
$\alpha Z$ and only terms of relative order $\alpha Z$ survive in
the Born matrix element. Terms of this order also
come from the continuum wave function.
Using exact wave functions for the continuum electron,
it turns out that the cross section calculated from plane
waves for the ejected electron is already wrong for small
$\alpha Z$.
It is therefore not
astonishing that Coulomb corrections can become very large even
at high energies.
For the derivation of the Sauter formula, nonrelativistic
bound state wave functions and approximate Sommerfeld-Maue wave
functions for the ejected electron were used.
Boyer \cite{Boyer} improved the calculation
by using exact bound state wave functions and Sommerfeld-Maue
wave functions for the continuum wave function.
The calculational advantage of using Sommerfeld-Maue wave functions
relies in the fact that the exact continuum wave functions
are only available as an expansion in partial waves, and
for higher energies summation over a large number of terms is
necessary in order to compute exact matrix elements.
But Sommerfeld-Maue wave functions are valid for the Coulomb
potential of point-like charges, and hence do not take into account
the finite size of the nucleus. We therefore pursue a similar
strategy, namely the eikonal distorted wave approximation,
where the advantage is that also finite size effects can be
taken into account. This is clearly necessary for nuclear
processes, whereas the relevant
physical length scale of the photoelectric
effect is given by $\sim (\alpha Z)^{-1} \lambda_C$ and much larger than
a typical nuclear radius.

In the following part of this paper, we revisit the eikonal
approximation by giving a condensed introduction to the
subject, where we also point out the most important facts
related to possible improvements concerning
the range of validity of the method in its basic form.
In sect. 3, we apply our calculation to coherent
vector meson electroproduction.
It turns out that the Coulomb corrections are indeed
large for heavy nuclei, and that it is necessary to use
a rather accurate description of the electrostatic
nuclear potential. We correct some errors in the literature
and rederive the eikonal result given in \cite{Kopeliovich}.

\section{Eikonal Approximation}
For highly relativistic particles in a potential $V$ with
asymptotic momentum $\vec{p}$,
one may neglect the mass of the
particle ($|E-V|, |\vec{p} \, | \gg m$),
such that the energy-momentum relation reduces to
($\hbar=c=1$)
\begin{equation}
(E-V)^2=\vec{p}^{\, 2} + m^2 \, \rightarrow \, E-V=p, \, p=|\vec{p} \,|.
\label{classical}
\end{equation}
The classical relation (\ref{classical}) for the (initial) momentum of the
particle $\vec{p}_i=p_i \hat{p}_i$ can be taken into account
approximately in quantum theory by modifying the plane wave
describing the initial state of the particle by the eikonal phase $\chi_1(\vec{r})$
\begin{equation}
e^{i\vec{p}_i\vec{r}} \, \rightarrow \, e^{i \vec{p}_i\vec{r}+i\chi_1(\vec{r})} \, ,
\end{equation}
\begin{equation}
\chi_1(\vec{r})=-\int \limits_{-\infty}^{0} V(\vec{r}+
\hat{p}_i s) ds \, .
\end{equation}
If we choose the initial momentum parallel to the $z$-axis
$\vec{p}_i=p_z^i \hat{{\bf{e}}}_z$,
the phase is
\begin{equation}
\chi_1(\vec{r})=-\int \limits_{-\infty}^{z} V(x,y,z') dz' \, .
\end{equation}
The $z$-component of the momentum then becomes
\begin{equation}
-i \partial_z e^{i p_z^i z+i\chi_1}=
(p_z^i-V)e^{i p_z^i z+i\chi_1},
\end{equation}
and further analysis shows that also the transverse momentum
modification is well approximated.
Analogously, for the final state wave function follows
\begin{equation}
e^{i \vec{p}_f\vec{r}-i\chi_2(\vec{r})},
\end{equation}
where
\begin{equation}
\chi_2(\vec{r})=-\int \limits_{0}^{\infty} V(\vec{r}+
\hat{p}_f s') ds' \, .
\end{equation}
For the sake of simplicity, we consider spinless particles first.
The spatial part of the free charged particle current
\begin{equation}
-ie[e^{-i \vec{p}_f\vec{r}} \vec{\nabla}
e^{i \vec{p}_i\vec{r}}-(\vec{\nabla}e^{-i \vec{p}_f\vec{r}})
e^{i \vec{p}_i\vec{r}}]= e(\vec{p}_i+\vec{p}_f)e^{i(\vec{p}_i-
\vec{p}_f)\vec{r}} 
\end{equation}
becomes
\begin{displaymath}
-ie[e^{-i \vec{p}_f\vec{r}+i\chi_2(\vec{r})} \vec{\nabla}
e^{i \vec{p}_i\vec{r}+i\chi_1(\vec{r})}-
(\vec{\nabla}e^{-i \vec{p}_f\vec{r}
+i\chi_2(\vec{r})}) e^{i \vec{p}_i\vec{r}+i\chi_1(\vec{r})}]=
\end{displaymath}
\begin{equation}
e(\vec{p}_i+\vec{p}_f+\vec{\nabla} \chi_1-\vec{\nabla} \chi_2)
e^{i(\vec{p}_i-\vec{p}_f)\vec{r}+i\chi (\vec{r})
} \, , \label{scalarcurr}
\end{equation}
where $e$ is the charge of the particle and $\chi(\vec{r})=
\chi_1(\vec{r})+\chi_2(\vec{r})$.
The spatial part of the current now contains the
additional eikonal phase, and the prefactor 
contains gradient
terms of the eikonal phase which represent essentially
the change of the electron momentum due to the attraction of
the electron by the nucleus.

So far we have only considered the modification of the
phase of the wave function, and for many problems,
like the one treated in this paper, this is a sufficient
approximation. It has been applied to elastic high energy
scattering of Dirac particles in an early paper by Baker \cite{Baker}.
However, the method leads, e.g. for quasielastic nucleon knockout
scattering of electrons on lead with
initial energy $\epsilon_1=300$ MeV and energy transfer $\omega=100$ MeV,
to errors up to 50\% in the calculated cross sections. 
The reason is that also the amplitude of
the incoming and outgoing particle
wave functions is changed due to the Coulomb attraction.
This fact is related to the classical observation
that an ensemble of negatively charged test particles approaching
a nucleus is focused due to its attractive potential.
For the sake of completeness, we give here a short discussion of this
effect.

Knoll \cite{Knoll} derived the focusing from a high energy
partial wave expansion, following previous results given
by Lenz and Rosenfelder \cite{Lenz,Rosenfelder}.
E.g., for the incoming particle wave in the vicinity of
the nucleus he obtained an approximation for the electron wave
function ($u_p$ is the constant free electron spinor)
\begin{equation}
\psi_i=e^{i \delta_+} (p'/p) e^{i \vec{p'}\vec{r}}
\{1+a_1r^2-2a_2\vec{p'}\vec{r}+ia_1r^2 \vec{p'}\vec{r}+ia_2[
(\vec{p'} \times \vec{r})^2+
\vec{\sigma} (\vec{p'} \times \vec{r})] \} u_p, \label{expknoll}
\end{equation}
where $\sigma$ describes spin changing effects.
A similar formula holds for the final state wave function.
The so-called effective momentum
$\vec{p'}$ is parallel to $\vec{p}$ and is given by the classical
momentum of the electron in the center of the nucleus, i.e.
$p'=|E-V(0)|$.
The parameters $a_{1,2}$ depend on the shape of the nuclear electrostatic
potential. For a homogeneously charged sphere with radius $R_s$
they are given by $a_1=-\alpha Z/6 p' R_s^3, \,
a_2=-3 \alpha Z/4 p'^2 R_s^2$.
The increase of the amplitude of the wave while passing
through the nucleus is given by the $-2a_2\vec{p'}\vec{r}$-term.
The $a_1 r^2$-Term accounts for the decrease of the focusing
also in transverse direction.
Imaginary terms like $ia_2
(\vec{p'} \times \vec{r})^2$ describe the deformation of the
wave front near the center of the nucleus. They could be obtained
correspondingly by an expansion of the eikonal phase in that region,
and $\delta_+$ is the eikonal phase in the center of the nucleus
$\chi(0)$.

Apart from the spin structure which is absent for scalar particles,
the eikonal approximation relies basically on the same strategy
for particles with or without spin, namely on the modification
of the wave function by the eikonal phase. But there are also
gradient terms present in the expression for the spinless current
(\ref{scalarcurr}), which lead to corrections to the current which 
are of comparable magnitude as those which result from the wave
focusing. Such terms are not explicitly present in the
Dirac expression for the current (\ref{current1}) below.
It is instructive to have a closer look at the
structure of the electron current in a potential.
A Gordon decomposition of the electron current
\begin{equation}
j^\mu=e \bar{\Psi} \gamma^\mu \Psi \label{current1}
\end{equation}
can be performed by using the Dirac equation
$[i \gamma^\mu (\partial_\mu+ieA_\mu)-m] \Psi=0$.
The current can be split into a
convective current and a spin current
\begin{displaymath}
j_G^\mu=\Biggl\{
\frac{ie}{2m} \bigl[ \bar{\Psi} \partial^\mu \Psi -(\partial^\mu
\bar{\Psi}) \Psi \bigl] -\frac{e^2}{m} \bar{\Psi}\Psi A^\mu \Biggr\}
\end{displaymath}
\begin{equation}
+\frac{e}{2m} \partial_\nu [\bar{\Psi} \sigma^{\mu \nu} \Psi ] \, ,
\label{current2}
\end{equation}
which are separately conserved and gauge invariant.
For exact solutions of the Dirac equation, the two forms
of the current are of course equivalent.
Here, the convective current has exactly the same structure as the
current of a scalar particle in a potential.
Kopeliovich {\em{et al.}} \cite{Kopeliovich}
calculate Coulomb corrections by describing the leptons as
spinless particles. The discussion above clarifies when
such an approximation is justified (see also sect. 3).

In the following, we will consider electrons with energies
of the order of several GeV. Then the impact of the focusing effect
on the matrix element will not be larger than about one percent even
for heavy nuclei like lead, where the electrostatic potential in the
center of the nucleus is ($Z=82$, $A=208$)
\begin{equation}
\sim \frac{3}{2} \frac{\alpha Z}{R_A} \sim 25 \,  \mbox{MeV},
\end{equation}
and $R_A \sim 7.1$ fm is the equivalent radius of a
homogeneously charged sphere, which we will call
"nuclear radius" for short in the following.
E.g., for a scattering process where the initial and final
momentum of the electron is of the order of $10$ GeV/c,
the focusing enters the cross section by a factor of the order
$(p'/p)^4=(10.025/10)^4 \sim 1.01$, according to (\ref{expknoll}).

\section{Coherent Electroproduction of Vector Mesons}
\subsection{Matrix Element}
Coherent electroproduction of vector mesons from virtual
photons plays an important role in the understanding of the
transition of soft diffractive models to quantum chromodynamics
\cite{Donnachie,Weise}. Models based on the
assumption of vector dominance have been applied successfully to
describe the data \cite{Yennie}.
Our intention is not to give a better model to describe the vector
meson production, but to explore the importance of Coulomb correction
effects. Therefore we use a schematic model, which captures some
essential features of the production process.
But it is clear that for a more realistic analysis a better
model should be used.

Our model is the one proposed in \cite{Kopeliovich},
which is inspired by the vector dominance model and
which allows to derive a relatively simple form for the
vector meson production amplitude on a nucleus
with mass number $A$ and nuclear charge $Z$.

We denote the energy momentum vectors of the initial and final electron
by $(\epsilon_{1,2},\vec{p}_{1,2})$ and
the momentum of the produced meson by
$\vec{p}_V$. $\vec{e}_V$ denotes the polarization vector of the meson.
The production amplitude is then given by
\begin{equation}
M(eA \rightarrow e'VA)= \int
\limits_{0}^{\infty} dx \, \vec{e}_V \cdot
\vec{f}(\vec{p}_1,\vec{p}_2,\vec{p}_V,x) ,
\end{equation}
where $\vec{f}=\vec{f}_1-\vec{f}_2$ and
\begin{displaymath}
\vec{f}_{1,2}(\vec{p}_1,\vec{p}_2,\vec{p}_V,x)=
\frac{1}{2 \omega_{1,2}} \frac{\partial}{\partial \omega_{1,2}}
\end{displaymath}
\begin{displaymath}
\times \int \frac{d^3 r}{r} \Bigl\{ \Bigl[ \epsilon_1 \vec{p}_2-\epsilon_2
\vec{p}_1 \Bigr] + \Bigl[ \epsilon_1 \vec{\nabla} \chi_2(\vec{r})
+\epsilon_2 \vec{\nabla} \chi_1(\vec{r}) \Bigr] \Bigr\}
\end{displaymath}
\begin{equation}
\times \exp[i \vec{\kappa} \vec{r} + i \chi_1(\vec{r})+i \chi_2
(\vec{r}) - i\omega_{1,2}r ]. \label{matrixel}
\end{equation}
For details concerning the derivation of eq. (\ref{matrixel}) we
refer to \cite{Kopeliovich} and the appendix of this paper.
We use a different sign convention for the eikonal phase than
\cite{Kopeliovich}.
The gradient terms are artefacts of the spinless treatment of
the electrons in \cite{Kopeliovich}
and therefore their physical significance in eq. (\ref{matrixel})
dubious at best.
However, from our discussion above follows that they can be neglected
for momentum transfer $Q^2=(\vec{p}_1-\vec{p}_2)^2-(\epsilon_1-\epsilon_2)
\gg R_A^{-2}$,
since they are of the order of $\alpha Z/R_A$, and
$| \epsilon_1 \vec{p}_2-\epsilon_2 \vec{p}_1 |= \sqrt{
\epsilon_1 \epsilon_2 Q^2}$.

In (\ref{matrixel}), we have
\begin{displaymath}
\omega_1^2=(1-x)(\epsilon_1-\epsilon_2)^2-x(1-x)
\vec{p}_V^{\, 2} -2x/B,
\end{displaymath}
\begin{equation}
\omega_2^2=(1-x)((\epsilon_1-\epsilon_2)^2-m_V^2)^2-x(1-x)
\vec{p}_V^{\, 2} -2x/B \label{omegadef}
\end{equation}
and $\vec{\kappa}=\vec{p}_1-\vec{p}_2-x \vec{p}_V$.
Note that in eq. (50) in \cite{Kopeliovich}, the squares for
$\omega_{1,2}$ are missing and the terms containing $B$ are wrong.
We give therefore a detailed derivation of eqns. (\ref{matrixel},
\ref{omegadef})
in the appendix, which is missing in \cite{Kopeliovich}.
Note also the different sign in front of $\omega_{1,2}$ in
the exponent. The misprints will be corrected in the electronic
preprint version of the paper \cite{Kopeliovich2}.

B is the slope parameter of the differential cross section.
For coherent electroproduction,
it is related to the mean charge nuclear radius squared
$\langle r^2_A \rangle_{rms}$ by (see e.g. \cite{Kopeliovich,Weise})
\begin{equation}
B=\frac{1}{3} \langle r^2_A \rangle_{rms} = \frac{1}{5} R_A^2.
\label{slope}
\end{equation}
At high energies and $\epsilon_{1,2} \gg \sqrt{Q^2}$,
the vectors $\vec{p}_1$, $\vec{p}_2$ and $\vec{p}_1-\vec{p}_2$ are
nearly parallel. Therefore we choose the $z$-axis along
$\vec{p}_1$ such that the vector $\vec{r}=(\vec{b},z)$ is given
by its $z$-component and the projection $\vec{b}$ to the
normal plane.
Using the relation
\begin{equation}
\int \limits_{-\infty}^{\infty} \frac{dz}{r} e^{i \vec{\kappa} \vec{r}-
i \omega r} = 2 K_0 \Bigl(b\sqrt{\kappa_L^2-\omega^2}  \Bigr), \quad
\label{besselformula}
\end{equation}
where $K_0$ is the modified Bessel function, and $\kappa_{L,T}$
the longitudinal and transverse components of $\vec{\kappa}$
with respect to $\vec{p}_1$, we obtain ($\chi=\chi_1+\chi_2$)
\begin{equation}
\vec{f}_{1,2}=\frac{\epsilon_1 \vec{p}_2-\epsilon_2 \vec{p}_1}
{\omega_{1,2}} \frac{\partial}{\partial \omega_{1,2}}
\int \limits_0^\infty db \int \limits_0^{2 \pi}
d \varphi K_0 \Bigl( b \sqrt{\kappa_L^2-\omega_{1,2}^2} \Bigr)
e^{i \kappa_T b \cos(\varphi)+i \chi(b)}
\end{equation}
Eq. (\ref{besselformula}) is valid also for
$\kappa_L^2-\omega^2<0$, whereas in
\cite{Kopeliovich} the sign in front of $\omega$ is problematic.
From
\begin{equation}
\int \limits_0^{2 \pi}
d \varphi e^{i \kappa_T b \cos(\varphi)}=2 \pi J_0
(\kappa_Tb), \quad \frac{\partial}{\partial z} K_0(z)=-K_1(z)
\end{equation}
follows
\begin{equation}
\vec{f}_{1,2}=\frac{2 \pi (\epsilon_1 \vec{p}_2-\epsilon_2 \vec{p}_1)}
{\sqrt{\kappa_L^2-\omega_{1,2}^2}}
\int \limits_0^\infty db \,  b  K_1
\Bigl(b \sqrt{\kappa_L^2-
\omega_{1,2}^2} \Bigr) J_0(\kappa_T b) e^{i \chi(b)}.
\end{equation}
Kopeliovich {\em{et al.}} \cite{Kopeliovich}
performed calculations for a model potential
\begin{equation}
V_{mono}(r)=-\frac{\alpha Z}{r} e^{-\lambda r} (1-e^{-\mu r}),
\label{pot1}
\end{equation}
with an infrared cutoff $\exp(-\lambda r)$, and the last
factor corresponds to the monopole form of the nuclear form factor
via
\begin{equation}
\mu^{-2}=\frac{\langle r^2_A \rangle_{rms}}{6} .
\end{equation}
The eikonal phase can be calculated analytically for such
a potential. One obtains
\begin{displaymath}
\chi(\vec{r})=\chi_1(\vec{r})+\chi_2(\vec{r})=
-\int \limits_{-\infty}^{\infty} dz \, V_{mono}(\vec{b},z)=
\end{displaymath}
\begin{equation}
2 \alpha Z \Bigl\{ K_0(\lambda b)-K_0[(\mu+\lambda)b] \Bigr\}.
\end{equation}

\begin{figure}[htb]
        \centering
        \includegraphics[width=12cm]{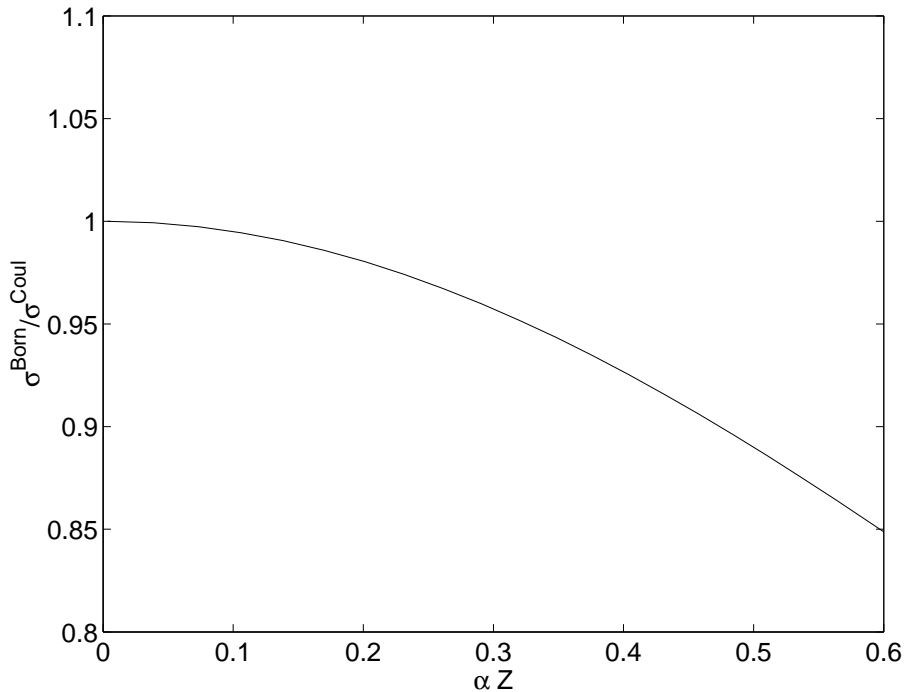}
        \caption{Coulomb correction for forward production
        for the model potential used in \cite{Kopeliovich},
        calculated for $\epsilon_1=100$ GeV,
        $y=(\epsilon_1-\epsilon_2)/\epsilon_1=0.6$, and
        electron scattering angle $1^o$ $(Q^2 \sim 1.2 \,
        \mbox{GeV}^2)$,
        with the slope parameter replaced by $B \rightarrow B/2$
        (see text).}
        \label{fig1}
\end{figure}

We checked the result given in Fig. \ref{fig4} in
\cite{Kopeliovich} for small $\lambda$.
Our results show the same behavior, although we obtain slightly
smaller Coulomb corrections. This might be due to the fact that
the authors used formula (\ref{slope}) only for their Sommerfeld-Maue
calculations \cite{Kopeliovich2}.
We can reproduce a nearly identical curve
(Fig. \ref{fig1}),
if we reduce the slope parameter $B$ according to formula
(\ref{slope}) by a factor of 2.

For the nuclear radius $R_A$
we used the formula
\begin{equation}
R_A=(1.128 \, \mbox{fm}) A^{1/3} + (2.24 \, \mbox{fm}) A^{-1/3},
\label{radiusrel}
\end{equation}
which is a good approximation for $A > 20$,
whereas the mass number and charge are related by
\begin{equation}
Z=\frac{A}{1.98+0.015 \, A^{2/3}}.
\end{equation}

\subsection{Improved Electrostatic Nuclear Potential}

It is obvious that the potential given by eq. (\ref{pot1}) provides
only an inaccurate description of the electrostatic nuclear potential
near the nucleus.
We looked therefore for potentials which allow an analytic calculation
of the eikonal phase ($\chi$ and $\chi_{1,2}$ as well).
This is not a trivial problem, since the class of
meaningful analytic potentials
which allow to express the eikonal phase by known special
functions is rather restricted.

A good choice is a potential energy for electrons of the form
\begin{equation}
(\alpha Z)^{-1}V_{model}(r)=-\frac{r^2+\frac{3}{2}R^2}{(r^2+R^2)^{3/2}}
-\frac{24}{25 \pi} \frac{R^2 R' r^4}{(r^2+R'^2)^4} \, ,
\label{pot_acc}
\end{equation}
which goes over into a Coulomb potential for $r \rightarrow
\infty$, and being close to the potential generated
by the relatively homogeneous spherical charge distribution of a nucleus.
The charge density
\begin{equation}
\rho(r)=-\frac{1}{e r} \partial^2_r (r V_{model}(r))
\end{equation}
corresponding to (\ref{pot_acc})
satisfies
\begin{equation}
\langle \rho \rangle = eZ \, ,
\end{equation}
\begin{equation}
\langle r^2 \rho \rangle = \frac{3}{5}R^2 eZ \, ,
\end{equation}
i.e. we can identify $R^2$ with the equivalent radius of the
homogeneously charged sphere which is given approximately by the formula
given above.
$R'$ serves as an additional fit parameter. A good choice is
$R'=0.5174 R$.

Because the eikonal phase turns out
to be divergent for a Coulomb-like potential,
we regularize the eikonal phase by subtracting a screening potential
$V_{scr} \sim (r^2+a^2)^{-1/2}$ from (\ref{pot_acc}),
such that the potential
falls off like $r^{-2}$ for large $r$.
The divergence can then be absorbed
in a constant divergent phase $\sim \log(a)$ without
physical significance, when the limit $a \rightarrow
\infty$ is taken. It is instructive to calculate the
eikonal phase for the simple screened potential
\begin{equation}
V(r)=-\frac{\alpha Z }{\sqrt{r^2+R^2}} \, , \quad
V^a(r)=-\frac{\alpha Z }{\sqrt{r^2+R^2}}+
\frac{\alpha Z }{\sqrt{r^2+a^2}} \, .
\end{equation}
One obtains for a particle incident parallel to the z-axis
with impact parameter $b$ ($r^2=b^2+z^2$)
\begin{displaymath}
\chi_1^a=\alpha Z \int_{-\infty}^{z} dz \,
\Bigl( \frac{1}{\sqrt{r^2+R^2}}-
\frac{1}{\sqrt{r^2+a^2}} \Bigr)=
\end{displaymath}
\begin{equation}
\alpha Z \log \frac{(z+\sqrt{z^2+b^2+R^2})(b^2+a^2)}
{(z+\sqrt{r^2+a^2})(b^2+R^2)},
\end{equation}
and therefore for the regularized
eikonal phase
\begin{equation}
\chi_1=\lim_{a \to \infty} (\chi_1^a - \alpha Z \log(a))=
\alpha Z \log \Bigl( \frac{z+\sqrt{r^2+R^2}}{b^2+R^2} \Bigr) .
\end{equation}

For a simple potential $\sim (r^2+R^2)^{-1/2}$, the
rms radius does not exist, since the corresponding
charge distribution does not fall off fast enough.

For the potential $V_{model}$ (\ref{pot_acc}), the regularized eikonal phase
is given by
\begin{equation}
(\alpha Z)^{-1} \chi(b)=\log \Bigl( \frac{R^2+b^2}{R^2} \Bigr)
-\frac{R^2}{R^2+b^2}
-\frac{3}{50}\frac{R^2 R_1 (R_1^4+4 R_1^2 b^2 + 8 b^4)}
{(R_1^2+b^2)^{7/2}}
\end{equation}
The $z$-dependent formulae for $\chi_{1,2}$ are a bit lengthy,
but can be derived in a straightforward manner.

\begin{figure}[htb]
        \centering
        \includegraphics[width=12cm]{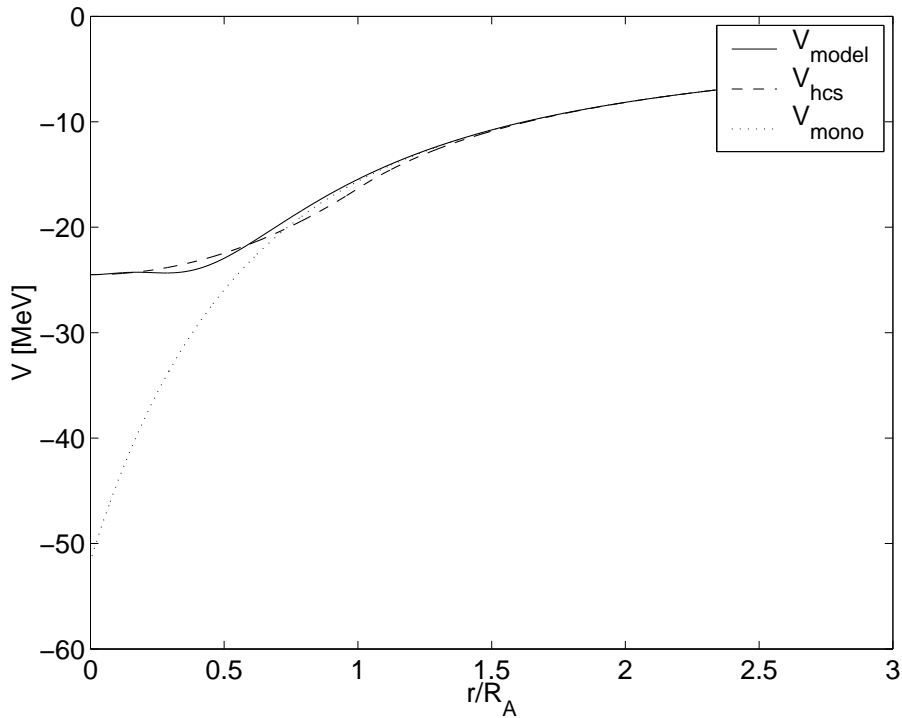}
        \caption{Comparison of different model potentials
        used in the calculations.}
        \label{fig2}
\end{figure}

In Fig. \ref{fig2}, we compare the potentials generated
by eq. (\ref{pot1}),
the potential $V_{hcs}$ of a homogeneously charged sphere with radius
$R_A$ and our model potential (\ref{pot_acc}) for
$^{208} Pb$ with identical mean squared radii in all three cases.

\section{Results and Conclusions}
\begin{figure}[htb]
        \centering
        \includegraphics[width=12cm]{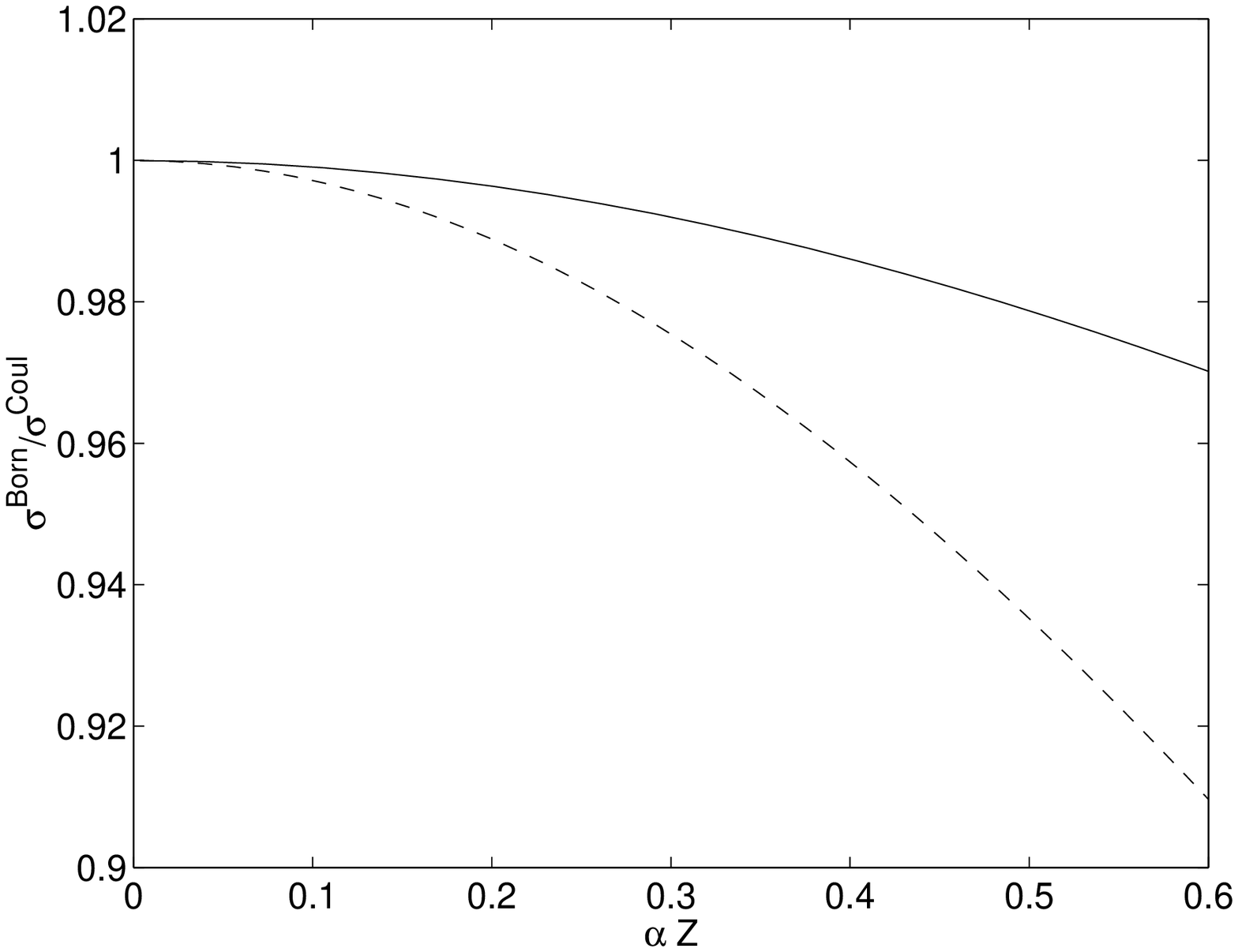}
        \caption{Comparison of Coulomb corrections for different
        potentials, given by eq. (\ref{pot_acc}), solid line,
        and eq. (\ref{pot1}), dashed line. Equal rms radii were
        used in both cases.}
        \label{fig3}
\end{figure}

For the results presented in Figs. 3-5, we used the same kinematic
conditions as for Fig. \ref{fig1}.
Fig. \ref{fig3} compares the eikonal correction, using the simple
model potential $V_{mono}$ and the potential $V_{model}$ given by
eq. (\ref{pot_acc}).
It turns out that the Coulomb corrections are overestimated
by the model potential given by eq. (23). This is probably due
to the fact that the potential is too deep in the
central nuclear region. It is therefore mandatory to use
an accurate description for the nuclear Coulomb potential
in order to obtain reliable results for the Coulomb corrections.

\begin{figure}[htb]
        \centering
        \includegraphics[width=12cm]{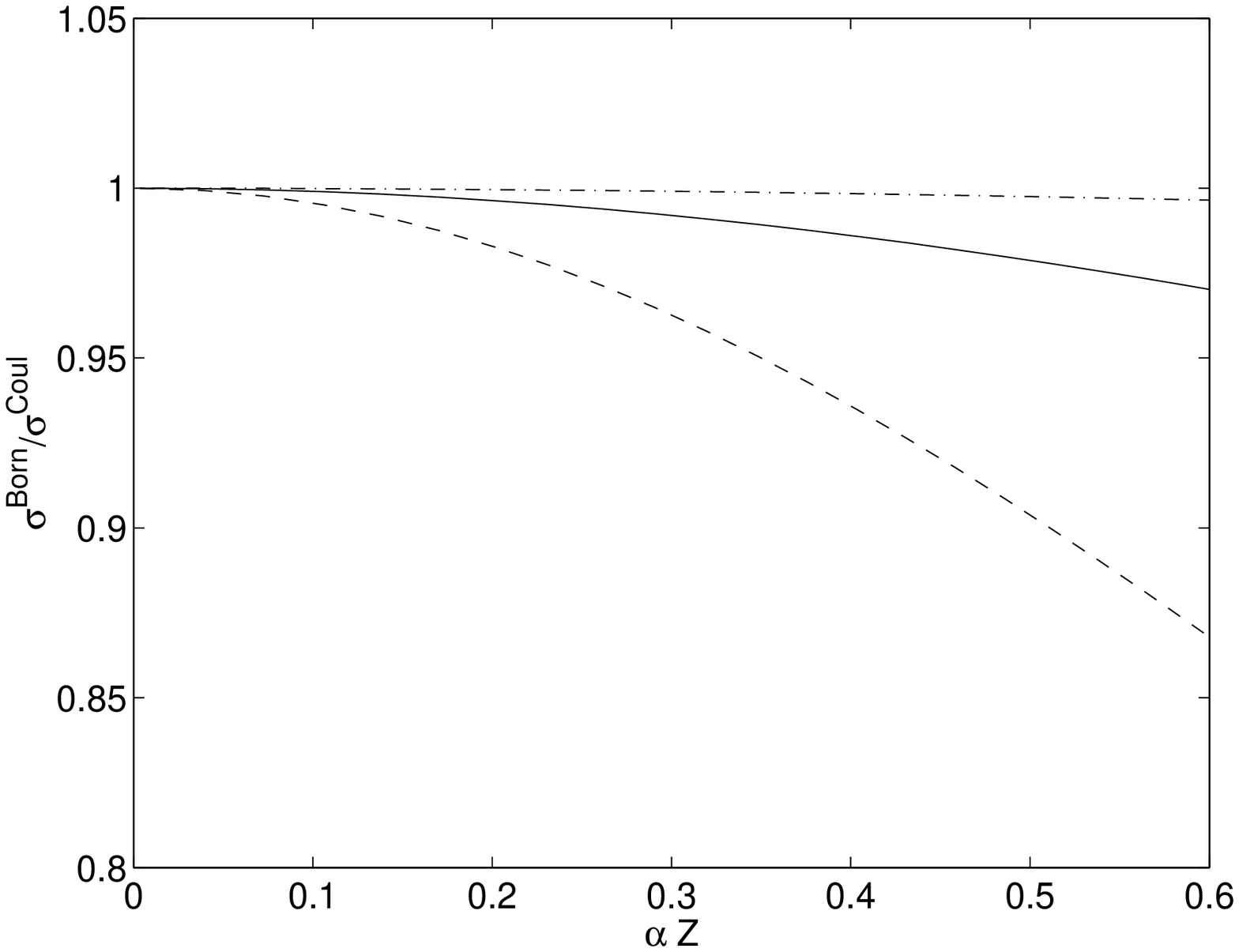}
        \caption{Coulomb corrections for the
        correct rms radius (solid line), doubled radius
        (dash-dotted), and radius divided by 2 (dashed).}
        \label{fig4}
\end{figure}

Fig. \ref{fig4} shows the Coulomb corrections for each element
with slope parameter $B$ according to eq. (\ref{slope}),
but with three different
rms charge radii for the electrostatic potential: the correct rms radius
$\sqrt \langle r^2_A \rangle_{rms}$,
a too large rms radius $2 \sqrt \langle r^2_A \rangle_{rms}$
and a too small rms radius $\frac{1}{2}\sqrt \langle r^2_A \rangle_{rms}$.
The figure clearly indicates that the distortion of the
electron waves is stronger for a small nucleus, whereas for
a larger nucleus, the eikonal phase varies less over the length
scale given by the nuclear radius (or slope parameter). The
initial and final state wave function resembles then more a plane
wave in the vicinity of the nucleus. The strong dependence of
the Coulomb corrections on the size of the nuclear radius clearly
indicates that the use of Sommerfeld-Maue wave functions
would lead to incorrect results.

\begin{figure}[htb]
        \centering
        \includegraphics[width=12cm]{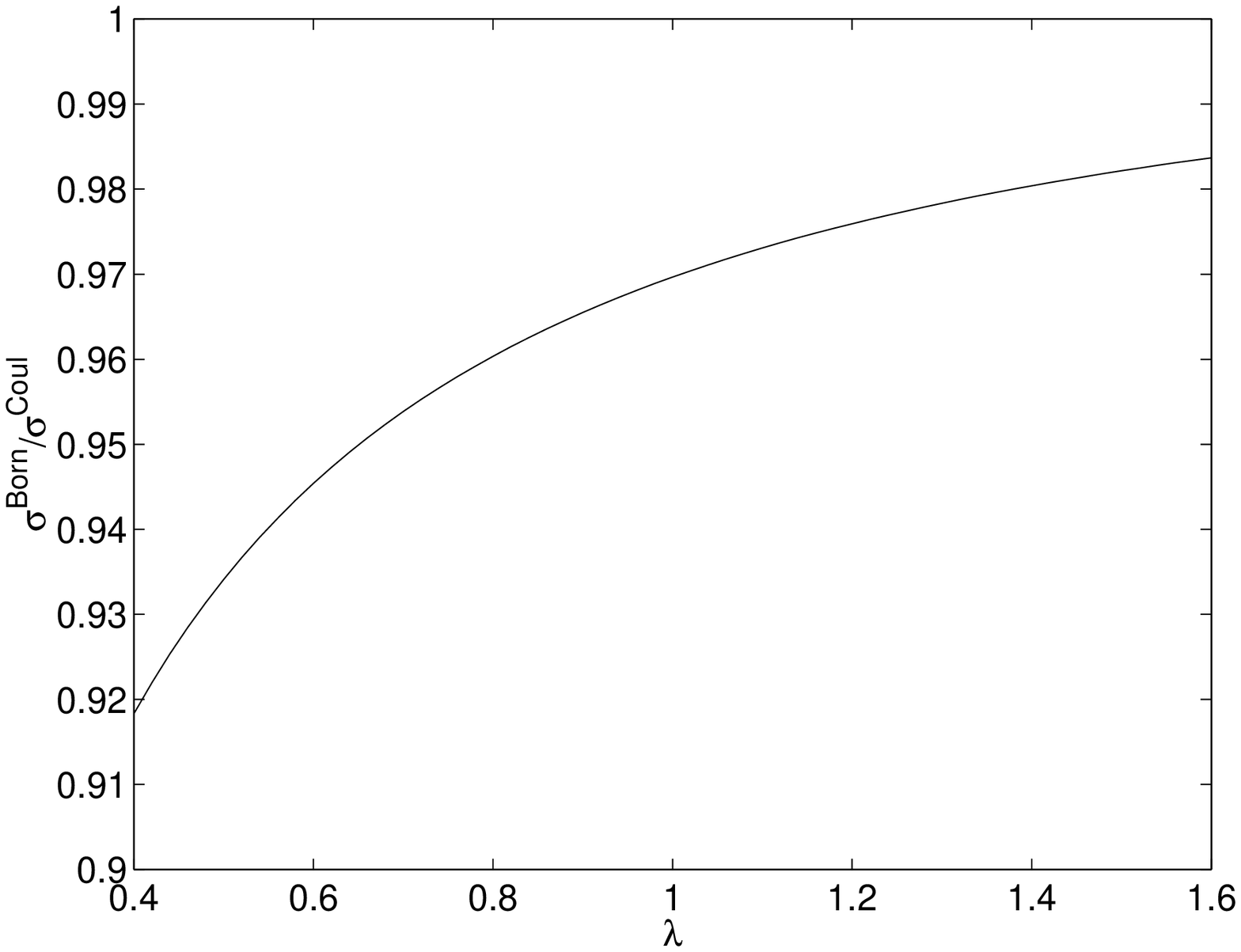}
        \caption{Coulomb corrections for different values $\lambda B$
        of the slope parameter for $^{208} Pb$.}
        \label{fig5}
\end{figure}

The dependence of the results on the slope parameter is displayed in
Fig. \ref{fig5} for $^{208} Pb$.
There we varied the slope parameter $B \rightarrow \lambda B$
for $\lambda=0.4 \ldots 1.6$, where $B$ is the theoretical value
given by eq. (\ref{slope}). The results show a clear dependence of the
Coulomb corrections on the model for the hadronic current.

\begin{figure}[htb]
        \centering
        \includegraphics[width=12cm]{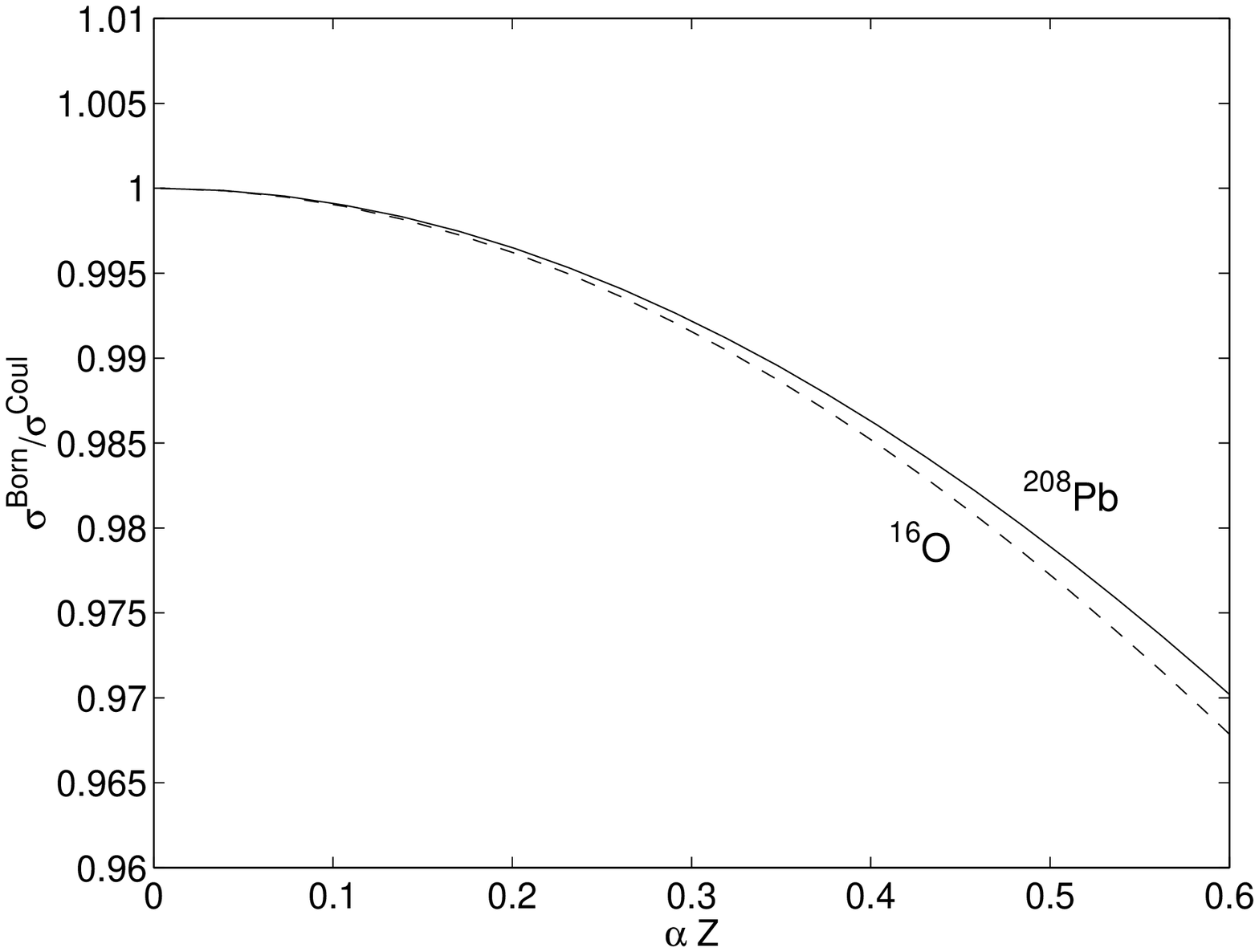}
        \caption{Dependence of Coulomb corrections on the
        nuclear charge, but with approximately correct nuclear radii
        according to eq. (\ref{radiusrel}) and correct
        slope parameter for lead and oxygen.}
        \label{fig6}
\end{figure}

Finally, Fig. \ref{fig6} shows the dependence of the Coulomb corrections
for $^{16}O$ and $^{208}Pb$, where we have varied artificially
the nuclear charge of the two elements,
while the correct charge radius of the two nuclei
and the corresponding slope parameter
were held fixed.

The calculations presented in this paper treat only the case
where the scattering angle of the electron is small, such that
the expression for the vector meson production amplitude
can be reduced to a simple two-dimensional integral, which
can be solved without involving large computational efforts.
But it is also possible to perform
numerical calculations of the three-dimensional integral
representation of the amplitude on a modern workstation,
such that arbitrary scattering angles and more general models
of the hadronic current could be treated. This is the subject of
a forthcoming paper.

The oscillatory behavior of the corrected cross section
shown in Fig. 3 in \cite{Kopeliovich} can not be reproduced by our
calculations. Fig. 4 shows clearly how important it is
to use the correct charge distribution of the nucleus for the
calculation of Coulomb corrections. Approximate wave functions
for pointlike nuclei are therefore not adequate.

\section{Appendix A. Matrix Element for Electroproduction
of Vector Mesons}

We start from the amplitude for coherent electroproduction of vector mesons
by electrons derived in \cite{Kopeliovich}, eq. (27), working with
the Born approximation first
\begin{equation}
M \propto \int \frac{d^3 q}{Q^2+i0} \Bigl[ \vec{j}(\vec{q})-j_0(\vec{q})
\frac{\vec{q}}{\nu} \Bigr] \vec{J}(Q,\vec{p}_V).
\end{equation}
where $\vec{j}$ is the spatial operator of the electron current,
$\vec{J}$ the hadronic current operator, $\nu=\epsilon_1-\epsilon_2$
and $Q^2=\vec{q}^{\, 2}-\nu^2$.
From current conservation
\begin{equation}
\nu j_0(\vec{q})=\vec{q} \cdot \vec{j} (\vec{q})
\end{equation}
we obtain
\begin{equation}
M \propto 2\frac{\epsilon_1 \vec{p}_2-\epsilon_2 \vec{p}_1}
{\epsilon_1-\epsilon_2} \int \frac{d^3q}{\vec{q}^{\, 2}-\nu^2}
\vec{J}(Q,\vec{p}_V).
\end{equation}
Using the hadronic model current ($\vec{\Delta}=\vec{q}-\vec{p}_V$ here)
\begin{equation}
\vec{J}(Q,\vec{p}_V)=\frac{\vec{e}_V m_V^2}{m_V^2+Q^2}
\frac{1}{1+B\vec{\Delta}^2/2}
\end{equation}
the amplitude becomes in Born approximation
\begin{equation}
M \propto 2\frac{(\epsilon_1 \vec{p}_2-\epsilon_2 \vec{p}_1)\vec{e}_V}
{\epsilon_1-\epsilon_2} 
\int d^3 r \int d^3 q
\frac{1}{(\vec{q}^{\, 2}-\nu^2)(\vec{q}^{\, 2}-\nu^2+m_V^2)}
\frac{(2/B) m_V^2}{(2/B+\vec{\Delta}^2)}
e^{i(\vec{p}_1-\vec{p}_2-\vec{q})\vec{r}} \, .
\end{equation}
Going over to the eikonal approximation, we modify the electron current
by the eikonal phases. Therefore, we have to evaluate the integral
\begin{equation}
I=
m_V^2 \int d^3 r \int
\frac{d^3 q}{(\vec{q}^{\, 2}-\nu^2)(\vec{q}^{\, 2}-\nu^2+m_V^2)
(2/B+\vec{\Delta}^2)}
e^{i(\vec{p}_1-\vec{p}_2-\vec{q})\vec{r}+i \chi_1(\vec{r})+i \chi_2(\vec{r})}.
\end{equation}
Due to the trivial identity
\begin{equation}
\frac{m_V^2}{Q^2 (Q^2+m_V^2)}=\frac{1}{Q^2}-\frac{1}{Q^2+m_V^2},
\end{equation}
we can decompose integral $I=I_1+I_2$ according to
\begin{displaymath}
I_1=\int d^3 r \int \frac{d^3 q}{(\vec{q}^{\, 2}-\nu^2)
(2/B+\vec{\Delta}^2)} e^{i(\vec{p}_1-\vec{p}_2-\vec{q})\vec{r}+i \chi_1(\vec{r})+i \chi_2(\vec{r})},
\end{displaymath}
\begin{equation}
I_2=\int d^3 r \int \frac{d^3 q}{(\vec{q}^{\, 2}-\nu^2+m_V^2)
(2/B+\vec{\Delta}^2)} e^{i(\vec{p}_1-\vec{p}_2-\vec{q})\vec{r}+i \chi_1(\vec{r})+i \chi_2(\vec{r})}.
\end{equation}
With Feynman's trick
\begin{equation}
\frac{1}{\alpha \beta}=\int \limits_{0}^{1}
\frac{dx}{[\alpha x + \beta (1-x)]^2}
\end{equation}
follows for $I_1$ first
\begin{displaymath}
I_1=\int \limits_{0}^{1} dx 
\int d^3 r \int d^3 q 
\frac{e^{i(\vec{p}_1-\vec{p}_2-\vec{q})\vec{r}+i \chi_1(\vec{r})+i \chi_2(\vec{r})}}
{[(\vec{q}^{\, 2}-\nu^2)(1-x)+(2/B+(\vec{q}-\vec{p}_V)^2)x]^2}=
\end{displaymath}
\begin{equation}
\int \limits_{0}^{1} dx 
\int d^3 r \int d^3 q 
\frac{e^{i(\vec{p}_1-\vec{p}_2-\vec{q})\vec{r}+i \chi_1(\vec{r})+i \chi_2(\vec{r})}}
{[(\vec{q}-x \vec{p}_V)^{\, 2}-x^2 \vec{p}_V^2-(1-x) \nu^2 +2x/B+x
\vec{p}_V^{\, 2}]^2}.
\end{equation}
Shifting the integration variable $\vec{q} \rightarrow
\vec{q}+x \vec{p}_V$
leads to
\begin{displaymath}
I_1=\int \limits_{0}^{1} dx 
\int d^3 r \int d^3 q 
\frac{e^{i(\vec{p}_1-\vec{p}_2-x\vec{p}_V-\vec{q})
\vec{r}+i \chi_1(\vec{r})+i \chi_2(\vec{r})}}
{[\vec{q}^{\, 2}-(1-x) \nu^2+x(1-x)\vec{p}_V^{\, 2} +2x/B]^2}=
\end{displaymath}
\begin{equation}
\frac{1}{2}\int \limits_{0}^{1} dx 
\int d^3 r \frac{1}{\omega_1} \frac{\partial}{\partial \omega_1}
\int d^3 q 
\frac{e^{i(\vec{p}_1-\vec{p}_2-x\vec{p}_V-\vec{q})
\vec{r}+i \chi_1(\vec{r})+i \chi_2(\vec{r})}}
{\vec{q}^{\, 2}-\omega_1^2}.
\end{equation}
Finally, the identity
\begin{equation}
\int d^3 q \frac{e^{-i \vec{q} \vec{r}}}{\vec{q}^{\, 2}-\omega_1^2}=
\pi \frac{e^{-i \omega_1 r}}{r}
\end{equation}
immediately gives the final result in agreement with eq.
(\ref{matrixel})
\begin{equation}
I_1=\int \limits_{0}^{1} dx 
\frac{\pi}{2 \omega_1} \frac{\partial}{\partial \omega_1}
\int \frac{d^3 r}{r} e^{i(\vec{p}_1-\vec{p}_2-x\vec{p}_V)\vec{r}
-i \omega_1 r+i \chi(\vec{r})}.
\end{equation}
For $I_2$ we must simply replace $\nu^2 \rightarrow \nu^2-m_V^2$.

\end{document}